\UseRawInputEncoding 
\documentclass[pra,onecolumn,superscriptaddress]{revtex4}%
\usepackage{amssymb}
\usepackage{amsmath}
\usepackage{graphicx}
\usepackage{dcolumn}
\usepackage{xcolor}
\usepackage{bm}
\usepackage{subfigure}
\usepackage{amsfonts}
\usepackage{appendix}
\usepackage{tikz}
\usetikzlibrary{quantikz2,backgrounds,fit,decorations.pathreplacing}
\usepackage{xcolor}%
\providecommand{\U}[1]{\protect\rule{.1in}{.1in}}

\begin{document}
\preprint{APS/123-QED}

\title{Learning  kernels with quantum optical circuits}

\author{ Aikaterini Mandilara}

\affiliation{Department of Informatics and Telecommunications, National and Kapodistrian
University of Athens, Panepistimiopolis, Ilisia, 15784, Greece}
\affiliation{Eulambia Advanced Technologies Ltd, Agiou Ioannou 24, Building Complex C, Ag. Paraskevi, 15342, Greece}

\author{ Aristeides D. Papadopoulos}

\affiliation{Eulambia Advanced Technologies Ltd, Agiou Ioannou 24, Building Complex C, Ag. Paraskevi, 15342, Greece}

\author{ Dimitris Syvridis}

\affiliation{Department of Informatics and Telecommunications, National and Kapodistrian
University of Athens, Panepistimiopolis, Ilisia, 15784, Greece}

\begin{abstract}

Support Vector Machines (SVMs) are a cornerstone of supervised learning, widely used for data classification. A central component of their success lies in kernel functions, which enable efficient computation of inner products in high-dimensional feature spaces. Recent years have seen growing interest in leveraging quantum circuits --both qubit-based and quantum optical-- for computing kernel matrices, with ongoing research exploring potential quantum advantages. In this work, we investigate two classical techniques for enhancing SVM performance through kernel learning --the Fisher criterion and quasi-conformal transformations-- and translate them into the framework of quantum optical circuits. Conversely, using the example of the displaced squeezed vacuum state, we demonstrate how established concepts from quantum optics can inspire novel perspectives and enhancements in SVM methodology. This cross-disciplinary approach highlights the potential of quantum optics to both inform and benefit from advances in machine learning.

\end{abstract}
\maketitle

\section{Introduction}

Support Vector Machines (SVMs) were originally introduced by V.~Vapnik and A.~Chervonenkis within the framework of statistical learning theory~\cite{Vapnik:1}, as a linear classification method based on maximizing the margin between data classes. The modern formulation, incorporating the soft margin and the kernel trick, was developed by C.~Cortes and V.~Vapnik in 1995~\cite{Vapnik:2}, enabling SVMs to handle non-linearly separable data through implicit mappings to high-dimensional feature spaces.

While neural networks have become the dominant approach for large-scale and high-dimensional classification tasks due to their expressiveness and scalability, SVMs continue to offer advantages in smaller-scale problems, particularly where interpretability, convex optimization, and theoretical guarantees are valued. Nonetheless, their applicability to large datasets is limited by computational constraints: training complexity scales at least quadratically with the number of samples, and kernel methods require the evaluation and storage of a full Gram matrix, which scales as $\mathcal{O}(n^2)$ in both time and space. Moreover, the performance of SVMs can be highly sensitive to the choice of kernel and hyperparameters, necessitating careful tuning—an effort that is often less pronounced in ensemble-based methods or deep learning architectures.

In 2019, the concept of quantum kernels was introduced~\cite{Schuld:1,Harrow:1}, along with the framework of Quantum Kernel Support Vector Machines (QKSVMs), marking a significant development in the application of quantum circuits and raising new questions within the field of Quantum Machine Learning (QML). A quantum kernel refers to a classical kernel function whose values are computed using quantum circuits. QKSVMs are a variant of SVMs in which the elements of the Gram matrix are evaluated quantum mechanically.

One immediate advantage of QKSVMs is the potential reduction in the computational and memory costs associated with evaluating and storing large Gram matrices, thereby addressing one of the major limitations of classical SVMs. A more subtle, yet potentially transformative, advantage may arise when quantum kernels are not only useful for solving a given learning task but also intractable to compute classically. Determining the conditions under which such quantum advantage emerges is an active area of research \cite{Thanasilp,Kubler,Ruslan,Florez}. The ongoing investigations—both those reporting positive results \cite{Liu,Huang,Jager} and those highlighting limitations—are expected to drive progress in both classical and quantum machine learning. 

In this work, we focus on a practical yet critical aspect of SVMs with kernels—and by extension, QKSVMs: the fine-tuning of hyperparameters within the kernel function. The importance of selecting the ``right'' kernel from a family of functions, and the performance benefits this choice can bring, have long been recognized in the classical machine learning community~\cite{Scholkopf}. This process, commonly referred to as \textit{kernel learning}, involves adapting the kernel to the training data and potentially incorporating additional criteria prior to applying the SVM.

Our interest lies in kernel learning techniques that are technically compatible with quantum circuits used for generating quantum kernels. Several recent works have begun to explore this direction, and we briefly review some of the key developments here. 
In her seminal work~\cite{schuld:2}, M.Schuld laid the foundation for interpreting variational quantum algorithms as kernel methods. Among other contributions, she examined how the choice of a quantum kernel influences the set of functions penalized in regularized empirical risk minimization problems~\cite{Scholkopf}. Moving toward practical implementations, the classical technique of \textit{train–target alignment} was employed in~\cite{Hubregtsen} to optimize hyperparameters in qubit-based quantum circuits. In simple terms, one constructs an ideal kernel Gram matrix  for the (labeled) training set and then tunes the kernel parameters to maximize alignment with this ideal matrix.
In their experiments with $5$-qubit circuits, this method led to a reduction in classification error of nearly $30\%$ compared to unoptimized SVM performance. A similar approach was applied in~\cite{Glick} using a superconducting qubit processor with $27$ qubits, generating quantum kernels tailored to data with group-theoretic structure—so-called \textit{covariant kernels}. Finally, the work in~\cite{Loyd:1} initiated the study of \textit{metric learning} in quantum circuits, a topic that, as we demonstrate in the present work, is closely connected to kernel learning and the SVM framework.

In this paper, we investigate the applicability of two classical kernel learning techniques within the context of quantum optical circuits: \textit{(a)} the Fisher criterion and \textit{(b)} quasi-conformal transformations  \cite{AMARI1999783, AMARI2002,Wu2007}. In the first case, we show that the Fisher criterion is closely related to \textit{quantum} metric learning based on the Hilbert--Schmidt distance~\cite{Loyd:1}, making it a natural and interchangeable tool in the quantum setting.
Quasi-conformal transformations, on the other hand, offer a more subtle approach to kernel learning and have seen somehow limited use in classical machine learning. Here, we demonstrate their potential benefits in the quantum regime and present specific scenarios where they enhance the performance of quantum kernel methods. While both techniques developed in this work are also applicable to qubit-based circuits~\cite{Tth}, we focus on continuous-variable systems, as they provide a more convenient framework for illustrating our results. In particular, quantum optical systems naturally incorporate Gaussian kernels, which are widely used in classical kernel methods.

The structure of this work is as follows. In Section~\ref{S2}, we review the foundational elements of SVMs with kernels, along with the concept of quantum kernels. This section also includes a brief overview of relevant literature on QKSVMs, with a focus on implementations using quantum optical circuits. Section~\ref{S3} forms the core of the paper, where we develop and present the theoretical framework for two kernel learning techniques adapted to quantum circuits.
In Section~\ref{S4}, we apply these techniques using kernels generated by the overlap of displaced squeezed vacuum states—referred to throughout as \textit{squeezed kernels}. These kernels are analytically tractable, experimentally accessible, and provide useful visualizations of the learning behavior. Moreover, squeezed kernels naturally incorporate tunable hyperparameters, making them well-suited for exploring the impact of kernel learning methods in a quantum setting.

\section{Kernels, SVMs with Kernels and quantum kernels \label{S2} }

In the present work, we address the problem of classification — a fundamental task in learning theory. We consider two classes of $n$-dimensional data: ${\mathbf{x}_i^{(A)}}$ for $i = 1, \ldots, N_A$, and ${\mathbf{x}_i^{(B)}}$ for $i = 1, \ldots, N_B$, with $N=N_A+N_B$. The goal is to classify a new, unseen data point. Assigning the label $y_i = 1$ to class $A$ and $y_i = -1$ to class $B$, the task becomes to determine the appropriate label $y$ for a given new input.
For simplicity, we assume that $\mathbf{x} \in \mathcal{X} \equiv \mathbb{R}^n$, and can be treated as a vector in a real vector space equipped with the standard inner product, $\left\langle\mathbf{x},\mathbf{x}' \right\rangle=\sum_j [\mathbf{x}]_j [\mathbf{x}']_j$,
where $[\mathbf{x}]_j$ denotes the $j$-th component (feature) of the vector $\mathbf{x}$.

A natural way to introduce the notion and use of kernels \cite{Scholkopf} is by viewing them as similarity measures between data vectors that aid in solving classification tasks. The simplest kernel function is the inner product,
$k(\mathbf{x},\mathbf{x}')=\left\langle\mathbf{x},\mathbf{x}' \right\rangle$, 
which assigns a real number to each pair of feature vectors and is symmetric under exchange of its inputs. However, such a simple similarity measure may not be sufficient to solve the task effectively. Fortunately, there exists a wide variety of more expressive kernel functions, which can be constructed by introducing a feature map that transforms the input vectors into a higher-dimensional (possibly infinite-dimensional) feature space:
\begin{eqnarray}
    \mathbf{\Phi}: & \mathcal{X}\rightarrow \mathcal{H}\\
                 & \mathbf{x}\rightarrow  \mathbf{\Phi}(\mathbf{x}),
\end{eqnarray}
where $\mathcal{H}$ is an inner product space (often called a Hilbert space). The kernel is then defined via the inner product in this feature space:
\begin{equation}
k(\mathbf{x},\mathbf{x}')=\left\langle \mathbf{\Phi}(\mathbf{x}),\mathbf{\Phi}(\mathbf{x}') \right\rangle \label{ke}
\end{equation}
which satisfies all the required properties of a valid kernel function \cite{Scholkopf}.

The method of SVMs \cite{Vapnik:1,Vapnik:2, Scholkopf} reveals an additional, important aspect of kernel functions: choosing a kernel corresponds to selecting a function space in which learning is performed. SVMs are hyperplane-based learning algorithms that can operate either in the original input space $\mathcal{X}$ or in a higher-dimensional feature space $\mathcal{H}$.
In this setting, the goal is to identify a hyperplane of the form $\left\langle \mathbf{w},\mathbf{\Phi}(\mathbf{x}) \right\rangle+b$, 
where $\mathbf{w} \in \mathcal{H}$ and $b \in \mathbb{R}$, that maximizes the margin of separation between the classes. The corresponding decision function is
\begin{equation}
    f(\mathbf{x})=sgn(\left\langle \mathbf{w},\mathbf{\Phi}(\mathbf{x}) \right\rangle+b). \label{hyper}
\end{equation}

Without going into the full derivation, the SVM solution is obtained by formulating and solving a dual optimization problem using Lagrange multipliers and the Karush–Kuhn–Tucker  conditions. This yields a decision function of the form:
\begin{equation}
   f(\mathbf{x})=sgn(\sum_i^m \alpha_i y_i\left\langle \mathbf{\Phi}(\mathbf{x}),\mathbf{\Phi}(\mathbf{x}_i) \right\rangle+b) \label{hyper2}  
\end{equation}
where $\alpha_i \geq 0$ and $b$ are determined by the solution of the dual problem, and the sum runs over the $m$ support vectors $\mathbf{x}_i$ — the training examples that lie closest to the decision boundary.

By applying the kernel identity from Eq.~(\ref{ke}), the decision function becomes:
\begin{equation}
   f(\mathbf{x})=sgn(\sum_i^m \alpha_i y_i k(\mathbf{x},\mathbf{x}_i)+b) \label{hyper3}  
\end{equation}
indicating that the classifier is expressed entirely in terms of the kernel function evaluated at the support vectors.

A widely used kernel is the Gaussian Radial Basis Function (RBF) kernel:
\begin{equation}
    k_{\gamma}(\mathbf{x},\mathbf{x}')=\exp{(-\frac{\gamma}{2} \left\| \mathbf{x}-\mathbf{x}'\right\|^2)} \label{RBF}
\end{equation}
where $\gamma > 0$ is a hyperparameter controlling the kernel width.
\vspace{1cm}

Recent advances in the realization of quantum circuits have motivated the introduction of quantum feature maps \cite{Schuld:1}, which refer to the encoding of classical data into quantum states using angle, amplitude, or Hamiltonian-based schemes. In this framework, a classical input $\mathbf{x}$ is mapped to a quantum state described by a density matrix:
\begin{equation}
\mathbf{x}\rightarrow \hat{\rho}_{\mathbf{x}}= \ket{ \mathbf{\Psi}_\mathbf{x}}\bra{ \mathbf{\Psi}_\mathbf{x}}. \label{qfm}   
\end{equation}
The corresponding quantum kernel is then defined as the Hilbert–Schmidt inner product between two such states:
\begin{equation}
    k\left(\mathbf{x},\mathbf{x}'\right)=\mathrm{Tr}\left[\hat{\rho}_{\mathbf{x}}\hat{\rho}_{\mathbf{x}'}  \right].\label{qk}
\end{equation}
In practice, the elements of the kernel (Gram) matrix $\mathbf{K}_{i,j}=\mathrm{Tr}\left[\hat{\rho}_{\mathbf{x}_i}\hat{\rho}_{\mathbf{x}_j}  \right]$ required by the SVM algorithm can be estimated using quantum subroutines such as the SWAP test or the inversion test \cite{Schuld:2021mml}.

 In the case where the encoding is implemented using an $N$-mode photonic quantum state—that is, in an infinite-dimensional Hilbert space, as in this work—the Wigner function formalism \cite{Schleich} provides a useful tool. The Wigner function of a state $\hat{\rho}_{\mathbf{x}}$ is defined as:
\begin{equation}
   W_{\mathbf{x}}\left(\mathbf{q},\mathbf{p}\right)= \frac{1}{\pi^{N}}\int \dots \int\bra{\mathbf{q}-\mathbf{y}}\hat{\rho}_{\mathbf{x}}\ket{\mathbf{q}+\mathbf{y}}e^{2i \mathbf{p}\cdot\mathbf{y}}d\mathbf{y}
\end{equation}
where $\mathbf{q}$ and $\mathbf{p}$ are the canonical quadrature variables.
In this setting, the quantum kernel can be expressed as an overlap integral of the corresponding Wigner functions over the $2N$-dimensional phase space:
\begin{equation}
    k\left(\mathbf{x},\mathbf{x}'\right)=(2\pi)^{N}\int \dots \int  W_{\mathbf{x}}\left(\mathbf{q},\mathbf{p}\right)W_{\mathbf{x}'}\left(\mathbf{q},\mathbf{p}\right)d\mathbf{q}d\mathbf{p}. \label{KW}
\end{equation}

Let us now review various encoding schemes into quantum optical states that have been considered in the literature. To begin, we introduce the single-mode displaced squeezed vacuum state, defined as:
\begin{equation}
    \ket{\alpha;\zeta} =\hat{D}(\alpha)\hat{S}(\zeta)\ket{0} 
   =  e^{\alpha \hat{a}^{\dagger}-\alpha^*\hat{a}}e^{\frac{1}{2}(\zeta\hat{a}^{\dagger2}-\zeta^*\hat{a}^2)}\ket{0}, \label{sd}
\end{equation}
where $\alpha=\frac{1}{\sqrt{2}}(q_0+i p_0)$ is the displacement in  phase-space,  $\zeta=re^{i\phi}$ is the complex squeezing parameter, 
and $\hat{a}^{\dagger}$ ($\hat{a}$) the creation (annihilation) operators of the mode. The state in Eq.~(\ref{sd}) depends on four real parameters:
$q_0 \in \mathbb{R},\ p_0 \in \mathbb{R},\ r \ge 0$, and $\phi \in [0, 2\pi)$.  

In the pioneering work \cite{Schuld:1}, where quantum kernels were first introduced, a family of kernel functions was studied based on encoding a single feature $[\mathbf{x}]_i$ into the angle $\phi$ of the squeezed vacuum state in Eq.~(\ref{sd}), while keeping $q_0 = p_0 = 0$ and treating the squeezing magnitude $r$ as a hyperparameter. To encode an entire input vector $\mathbf{x}$, this approach requires $n$ independent modes, resulting in a product state of the form: $\ket{\alpha=0;\phi=[\mathbf{x}]_1}\otimes\ldots\ket{\alpha=0;\phi=[\mathbf{x}]_n}$. In a subsequent study \cite{LI2022128088}, the so-called squeezing amplitude encoding was proposed, where each feature is mapped as $[\mathbf{x}]_i \mapsto r_i$, with $\alpha = 0$.  Finally, in \cite{MEHTA2024129704}, the encoding $[\mathbf{x}]_i \mapsto q_0$ was considered, corresponding to displacement along the position quadrature of a displaced squeezed vacuum state.

In all of the encoding schemes mentioned above, the corresponding kernel functions can be analytically evaluated using Eq.~(\ref{KW}) and Gaussian integration or other tractable methods. As a result, these kernels are not generally expected to offer a significant advantage over classical kernel functions.
However, if Gaussian encodings are followed by non-Gaussian operations, as proposed in \cite{Schuld:1, Harrow:1}, the resulting kernel functions may become classically intractable—especially in high-dimensional settings, where multimode optical states must be considered.
In the present work, we demonstrate the proposed learning scheme using examples in which encoding is performed via the parameters of the displaced squeezed vacuum state Eq.~(\ref{sd}). Our goal is to illustrate an approach that could be extended to more complex scenarios, where the elements of the Gram matrix may only be accessible through quantum optical experiments.

\section{Learning quantum kernels \label{S3}}

Suppose now that a quantum kernel is to be used within the SVM framework, implementing the so-called QKSVM method. One immediately encounters the same fundamental challenge as in the classical case: selecting or adjusting the hyperparameters of the quantum kernel—i.e., the settings of the quantum circuit—to match the geometry and structure of the dataset under study. 
It is therefore natural to seek inspiration from established classical methods for kernel learning, and to reinterpret these approaches in a quantum context. The strategy we adopt is similar in spirit to that proposed in \cite{Hubregtsen}, where the train–target alignment method was applied to quantum kernels derived from qubit-based circuits. 
In the present work, we explore two independent classical strategies for kernel learning and adapt them in Section~\ref{S4} to our photonic quantum setting: \textit{(a)} the Fisher criterion and \textit{(b)} quasi-conformal transformations method.

\subsection{Fisher criterion and quantum metric learning with Hilbert-Schmidt distance}

Fisher criterion-based kernel learning involves optimizing a kernel function within the framework of Kernel Fisher Discriminant Analysis (KFDA) to maximize class separability. This is achieved by increasing the ratio of between-class scatter to within-class scatter.
If we omit the final step of extracting nonlinear discriminant features and instead focus on selecting among a family of kernel functions parameterized by a hyperparameter, the Fisher criterion provides a principled way to choose that hyperparameter. Specifically, it selects the value that maximizes the distance between class means while minimizing the variance within each class in the feature space defined by the kernel.

We now formulate the Fisher criterion in the context of the quantum feature map defined in Eq.(\ref{qfm}), applied to the binary classification problem introduced at the beginning of Section\ref{S2}.
To this end, we define the vector of quantum-mapped data points corresponding to the input vectors $\mathbf{x}$ as:
\begin{equation}
\mathbf{\hat{P}(x)}=\left(\hat{\rho}_{\mathbf{x}_1},\ldots,\hat{\rho}_{\mathbf{x}_N}\right)
\end{equation}
where each quantum state satisfies $\mathrm{Tr}(\hat{\rho}^2_{\mathbf{x}}) = 1$.
We further distinguish the states corresponding to each class, labeling them as $\hat{\rho}^{(A)}$ and $\hat{\rho}^{(B)}$. The associated vectors of quantum states for each class are defined as:
\begin{eqnarray}
\mathbf{\hat{P}^A(x)}& =\left(\hat{\rho}^{(A)}_{\mathbf{x}_1},\ldots,\hat{\rho}^{(A)}_{\mathbf{x}_{N_A}}\right) \nonumber\\
\mathbf{\hat{P}^B(x)}& =\left(\hat{\rho}^{(B)}_{\mathbf{x}_1},\ldots,\hat{\rho}^{(B)}_{\mathbf{x}_{N_B}}\right).
\end{eqnarray}
Finally, we define the average density matrices for each class, constructed from the corresponding ensembles of quantum states:
\begin{equation}
    \hat{\rho}_{A}=\frac{1}{N_A}\sum_{i=1}^{N_A}\hat{\rho}^{(A)}_{\mathbf{x}_i}, ~~\hat{\rho}_{B}=\frac{1}{N_B}\sum_{i=1}^{N_B}\hat{\rho}^{(B)}_{\mathbf{x}_i}.\label{rhoAB}
\end{equation}

The total scatter matrix $\mathbf{S}_T$ of the mapped data in the quantum feature space is defined as:
\begin{equation}
    \mathbf{S}_T=\frac{1}{N} \mathbf{\hat{P}(x)}\left(\mathbb{\hat{I}}-\mathbf{\hat{1}}_{N\times N}\right)\mathbf{\hat{P}^{T}(x)}
\end{equation}
where $\hat{\mathbb{I}}$ denotes the $N \times N$ identity matrix, and $\hat{\mathbf{1}}_{N \times N}$ is the $N \times N$ matrix with all entries equal to $1/N$.

The within-class, or intraclass, scatter matrix of the mapped data is defined as:
\begin{equation}
    \mathbf{S}_I=\frac{1}{N_A} \mathbf{\hat{P}}^A\mathbf{(x))}\left(\mathbb{\hat{I}}-\mathbf{\hat{1}}_{N_A\times N_A}\right)\mathbf{\hat{P}}^{A^T}\mathbf{(x)}+\frac{1}{N_B} \mathbf{\hat{P}}^B\mathbf{(x))}\left(\mathbb{\hat{I}}-\mathbf{\hat{1}}_{N_B\times N_B}\right)\mathbf{\hat{P}}^{B^T}\mathbf{(x)}.
\end{equation}
A straightforward calculation leads to the identity: 
\begin{equation}
    \mathrm{Tr}(\mathbf{S}_I)=2-\mathrm{Tr}( \hat{\rho}_{A}^2)-\mathrm{Tr}( \hat{\rho}_{B}^2),\label{SI}
\end{equation}
which admits a clear quantum mechanical interpretation:
for each class, the more concentrated (less scattered) the mapped quantum states are in Hilbert space, the higher the purity $\mathrm{Tr}(\hat{\rho}^2)$ of the corresponding ensemble average, and thus the lower the intraclass scatter $\mathrm{Tr}(\mathbf{S}_I)$.
Accordingly, a well-chosen kernel should minimize this quantity, thereby compressing the variance within each class in the feature space.

At the same time, it is important to ensure that the interclass distances are large in the quantum feature space. To quantify this aspect, we define the between-class, or interclass, scatter matrix as:
\begin{equation}
    \mathbf{S}_O=\mathbf{S}_T-\mathbf{S}_I.
\end{equation}
In the special case where $N_A=N_B=n/2$, algebraic manipulations lead to the following compact expression:
\begin{equation}
    \mathrm{Tr}(\mathbf{S}_O)=\frac{3}{4} \mathrm{Tr}( \hat{\rho}_{A}^2)+\frac{3}{4} \mathrm{Tr}( \hat{\rho}_{B}^2)-
    \frac{1}{2} \mathrm{Tr}( \hat{\rho}_{A}  \hat{\rho}_{B})-1. \label{SO}
\end{equation}
As expected, the overlap between the two ensembles of quantum states, given by $\mathrm{Tr}( \hat{\rho}_{A}  \hat{\rho}_{B})$,  appears in the expression. The smaller this overlap, the larger the value of 
 $\mathrm{Tr}(\mathbf{S}_O)$, indicating greater separation between the two classes in the quantum feature space. Therefore, maximizing 
 $\mathrm{Tr}(\mathbf{S}_O)$ becomes a desirable objective when selecting or tuning the quantum kernel.

Using Eqs.~(\ref{SI}) and (\ref{SO}), we arrive at the following expression for the Fisher score:
\begin{equation}
    Fs=\frac{\mathrm{Tr}(\mathbf{S}_O)}{\mathrm{Tr}(\mathbf{S}_I)}= \frac{1-
    \mathrm{Tr}( \hat{\rho}_{A}  \hat{\rho}_{B})}{2 \left(2-\mathrm{Tr}( \hat{\rho}_{A}^2)-\mathrm{Tr}( \hat{\rho}_{B}^2)\right)}-\frac{3}{4}. \label{Fs}
\end{equation}
Maximizing this ratio promotes both high between-class separability (numerator) and low within-class variance (denominator), making it a natural objective for kernel selection. This formulation provides a criterion for identifying an optimal quantum kernel within a given family, such as those parametrized by encoding or circuit settings. In the case of imbalanced datasets, where 
$N_A\ne N_B$ the expression in Eq.~(\ref{Fs}) becomes slightly more involved, but the underlying principle remains the same.

Interestingly, a similar expression to  the Fisher criterion, Eq.(\ref{Fs}), has  been derived via an alternative approach in \cite{Loyd:1}, where quantum metric learning is explored. That work also provides detailed methods for experimentally estimating the quantities appearing in Eq.(\ref{Fs}) within qubit-based quantum circuits. The extension of these techniques to quantum optical circuits is straightforward. While the Fisher criterion serves as a preparatory step for applying SVMs with kernels—hence also for QKSVMs—quantum metric learning has not yet been employed for this purpose. In what follows, we briefly review the relevant results from \cite{Loyd:1}, particularly concerning the Hilbert–Schmidt distance metric, and establish the connection to QKSVM that has so far remained unaddressed.

In brief, the core idea presented in \cite{Loyd:1} is as follows: adjust the parameters of a variational quantum circuit to maximize the Hilbert–Schmidt distance between the ensembles defined by the two data sets, as given in Eq.~(\ref{rhoAB}). Classification of test data is then performed by evaluating the fidelity between the test instance and the learned class distributions. In our context, we propose a natural substitution: replace the family of variational quantum circuits with a family of quantum kernels, and interpret the tunable weights of the circuit as the hyperparameters of the quantum kernels.

In more detail, \cite{Loyd:1} introduces a fidelity-based classifier to construct the decision function for a test state $\hat{\rho}_{\mathbf{x}}$, defined as:
\begin{equation}
    f_{fid}(\mathbf{x})= \mathrm{sgn}(Tr\left[ \hat{\rho}_{\mathbf{x}}\left(\hat{\rho}_{A}-\hat{\rho}_{B}\right)\right]).\label{fid}
\end{equation}
When a linear loss function is considered, the corresponding empirical risk $\hat{I}\left[f_{fid}\right]$ to be minimized is given by $- D_{HS}(\hat{\rho}_A,\hat{\rho}_B)$, where $D_{HS}$ denotes the Hilbert–Schmidt distance between the two density operators, defined as:
\begin{equation}
    D_{HS}(\hat{\rho}_A,\hat{\rho}_B)= Tr\hat{\rho}_A^2+Tr\hat{\rho}_B^2-2 Tr\hat{\rho}_A \hat{\rho}_B. \label{dhs}
\end{equation}
In summary, the hyperparameters of the quantum kernel (or, equivalently, the weights in the variational quantum circuit) are optimized to maximize the Hilbert–Schmidt distance $D_{HS}$, as defined in Eq.(\ref{dhs}), between the class-conditional ensembles derived from the training data. Classification of a new data point, represented by the state $\hat{\rho}_{\mathbf{x}}$, is then performed using the fidelity-based decision function in Eq.(\ref{fid}).

The similarity between Eqs.(\ref{Fs}) and (\ref{dhs}) is evident—both criteria yield equivalent results, particularly when an additional hyperparameter is introduced in the overlap term of Eq.(\ref{dhs}), as investigated in \cite{Rth}. Moreover, one can observe that the application of the QKSVM method naturally follows quantum metric learning based on the Hilbert–Schmidt distance. In other words, quantum metric learning can be interpreted as a kernel learning procedure.

To illustrate this, we decompose the decision function in Eq.(\ref{fid}) using the individual components of the class-conditional density operators defined in Eq.(\ref{rhoAB}):
\begin{equation}
    f_{fid}(\mathbf{x})=\mathrm{sgn}(\sum_i^{N_A} \frac{1}{N_A}k(\mathbf{x},\mathbf{x}_i^{(A)})-\sum_i^{N_B} \frac{1}{N_B}k(\mathbf{x},\mathbf{x}_i^{(B)})) \label{fd}
\end{equation}
where we have used the definition of the quantum kernel, Eq.~(\ref{qk}).
 This formulation reveals that maximizing $D_{HS}$  can be viewed as optimizing the kernel function $k(\mathbf{x},\mathbf{x}')$ to improve the accuracy of the decision function on the training data.Once the kernel has been learned, it can be passed to an SVM, enhancing the fidelity-based decision function in Eq.~(\ref{fd}). Specifically, the uniform averages over the training sets are replaced by learned weights (dual variables) from the SVM optimization:
\begin{equation}
    f_{SVM}(\mathbf{x})=\mathrm{sgn}(\sum_i^{N_A} \alpha_i k(\mathbf{x},\mathbf{x}_i^{(A)})-\sum_i^{N_B} \beta_i k(\mathbf{x},\mathbf{x}_i^{(B)})+\alpha_0)
\end{equation}
where the nonzero weights 
$\alpha_i,~\beta_i\ge0$ correspond to support vectors.

From a quantum perspective, this second step of applying SVM can be interpreted as reconstructing the class-conditional density matrices in Eq.~(\ref{rhoAB}) with modified contributions: the support vectors are emphasized via their associated weights, while the remaining training examples are effectively down-weighted or eliminated. This re-weighting leads to maximization of the margin between the two classes in the quantum feature space.

In this work, we adopt a kernel learning approach by maximizing the Hilbert–Schmidt distance, Eq.(\ref{dhs})—the quantum analogue of the classical Fisher criterion—and subsequently apply the QKSVM method. Our example demonstrates that this approach can achieve superior results compared to the simpler strategy of maximizing the Hilbert–Schmidt distance followed by classification using the fidelity-based decision function, Eq.(\ref{fid}), as proposed in \cite{Loyd:1}.

\subsection{Quasi-conformal transformations on quantum kernels}

Let us briefly describe the idea introduced by S. Wu and S. Amari \cite{AMARI1999783, AMARI2002} of applying conformal transformations to kernel functions, with the aim of improving the performance of SVMs. The approach begins with an initial kernel function $k(\mathbf{x},\mathbf{x}')$ , which is used in a standard kernel SVM to identify the support vectors. In a second stage, a modified kernel $\tilde{k}(\mathbf{x},\mathbf{x}')$ is constructed by applying quasi-conformal transformations to the original kernel. This transformation is designed to magnify regions around the support vectors identified in the first stage, thereby refining the focus of the SVM on critical regions of the input space.

A key notion for describing transformations on input vectors—such as those induced by (quantum) feature maps—is the Riemannian metric $ g_{ij}$, associated with the transformation. While this metric can be evaluated using the Jacobian of the transformation, it is more convenient for our purposes to use the following formula, which directly relates the metric to the kernel function of the transformation \cite{AMARI1999783}:
\begin{equation}
    g_{ij}(\mathbf{x})=\left.\frac{\partial}{\partial x_i}\frac{\partial}{\partial x_j} k(\mathbf{x},\mathbf{x}')\right|_{\mathbf{x}'=\mathbf{x}}. \label{g}
\end{equation}
From the Riemannian metric, one can infer various properties of the transformation. Here, we focus on two key quantities: the \textit{magnification factor}, defined as $\sqrt{G(\mathbf{x})}=\sqrt{Det[g_{ij}(\mathbf{x})]}$, which characterizes local changes in spatial geometry; and the principal spread direction, given by the eigenvector of $ g_{ij}$ corresponding to its largest eigenvalue.

A transformation is called conformal if it modifies the Riemannian metric as 
\begin{equation}    
\tilde{g}_{ij}(\mathbf{x})=\Omega(\mathbf{x})g_{ij}(\mathbf{x}).
\end{equation}
where $\Omega(\mathbf{x})$
is a positive scalar function. Conformal transformations preserve angles throughout the input space, but induce local magnification factors of $\sqrt{\Omega(\mathbf{x})}$.

The goal in our context is to apply an additional conformal transformation to the initial feature map (or, equivalently, to the kernel function) used in SVMs, such that the resulting kernel emphasizes regions around the support vectors. However, designing data-dependent conformal transformations is generally difficult, as will be illustrated in the example in the next section. To address this challenge, the authors in \cite{AMARI1999783} proposed the use of quasi-conformal transformations, obtained by modifying the original kernel as:
\begin{equation}
        \tilde{k}(\mathbf{x},\mathbf{x}')=q(\mathbf{x})q(\mathbf{x}')k(\mathbf{x},\mathbf{x}') \label{qct}
    \end{equation}
    where $q(\mathbf{x})$ is a positive scalar function.
Quasi-conformal transformations preserve the local structure of the space to some extent, although they may not strictly preserve angles everywhere \cite{Scholkopf}. The corresponding Riemannian metric is modified as:
    \begin{equation}
        \tilde{g}_{ij}(\mathbf{x}) = \frac{\partial q(\mathbf{x})}{\partial x_i} \frac{\partial q(\mathbf{x})}{\partial x_j} + q(\mathbf{x})^2 g_{ij}(\mathbf{x}), \label{til}
    \end{equation}
which implies that the local magnification factor now depends explicitly on the choice of $q(\mathbf{x})$.

While it is not immediately clear how to implement conformal transformations directly within quantum circuits that produce kernel functions, quasi-conformal transformations can be straightforwardly realized in both qubit-based and quantum optical circuits. In the qubit setting, one can simply append additional qubits initialized in the ground state $\ket{\mathbf{0}}$, apply a data-dependent unitary transformation $\hat{U}(\mathbf{x})$ and estimate the probability $\mathcal{P}(0)=\left|\bra{\mathbf{0}}\hat{U}(\mathbf{x})\ket{\mathbf{0}}\right|^2$ of measuring all qubits in the ground state. This probability defines a data-dependent scaling factor $q(\mathbf{x})$ in the modified kernel of Eq.~\eqref{qct}, given by:
 \begin{equation}
   q(\mathbf{x})= \mathcal{P}(0)= \bra{\mathbf{0}}\hat{U}(\mathbf{x})\ket{\mathbf{0}}\bra{\mathbf{0}}\hat{U}^{\dagger}(\mathbf{x})\ket{\mathbf{0}}.
 \end{equation}
A similar approach can be applied in quantum optical circuits, where one introduces additional, unentangled modes into the system. These modes can be acted upon by a data-dependent transformation, and the probability of measuring them in their vacuum (ground) state again defines the function $q(\mathbf{x})$.
 
Up to this point, the general idea of the method is clear. We now turn to its most subtle aspect: the design of the scaling factor $q(\mathbf{x})$, such that the modified kernel leads to improved performance in the second application of the SVM. In the foundational works \cite{AMARI1999783, AMARI2002, Wu2007, Qiuze2015}, both the theoretical developments and empirical demonstrations were carried out using the RBF kernel $ k_{\gamma}(\mathbf{x},\mathbf{x}')$, defined in Eq.~\eqref{RBF}. To make the discussion concrete, we will summarize the proposed strategies for engineering $q(\mathbf{x})$ using this kernel as a guiding example.

In the first run of the SVM, the standard RBF kernel $ k_{\gamma}(\mathbf{x},\mathbf{x}')$ is used, with the parameter $\gamma$ chosen according to the data scale --usually as the standard deviation of data. This yields a set of $m$ support vectors, denoted
 $\mathbf{x}^s_j$. The scaling factor $q(\mathbf{x})$ then constructed as a sum of RBF kernels centered at these support vectors:
\begin{equation}
  q(\mathbf{x})=\sum_j^m h_j  k_{\tilde{\gamma}_j}(\mathbf{x}^s_j,\mathbf{x}).  \label{sum}
\end{equation}

In the original work introducing the method \cite{AMARI1999783}, the authors set all
$\tilde{\gamma}_j$ to a uniform value chosen based on the resolution needed in the second SVM run. Since the goal is to magnify regions around support vectors, it is natural to choose $\tilde{\gamma}\ge \gamma$. Numerical experiments in both \cite{AMARI1999783} and our own studies \cite{Tth} suggest that $\tilde{\gamma}\approx 4~\gamma$ provides near-optimal performance. In the same work, the coefficients $h_j$   were set to the dual weights $\alpha_j$ obtained from the first SVM, as defined in Eq.~\eqref{hyper2}.

In a subsequent study \cite{AMARI2002}, the authors explored a different strategy to enhance performance: they set the coefficients $h_j$ to a uniform value and introduced a more refined, data-dependent scheme for selecting $\tilde{\gamma}_j$. This method incorporates local geometric information through a simple search in the input space, and was shown to improve results. Later, in \cite{Wu2007}, a different formulation of $q(\mathbf{x})$, not based on Eq.\eqref{sum}, was investigated. However, according to our numerical investigations, the most effective approach is that of \cite{Qiuze2015}, where Eq.\eqref{sum} is retained with a uniform $\tilde{\gamma}_j$,  while the weights $h_j$ are identified by the application of Fisher's criterion. Nonetheless, in this case, the improved performance obviously stems from  both Fisher-based feature weighting and the conformal transformation framework itself.

Based on the above partial review of the classical kernel literature, it becomes clear that there is no universal recipe for selecting the quasi-conformal factors $q(\mathbf{x})$. Our own numerical experiments further support this observation: the effectiveness of the method appears to be highly case-dependent, with the greatest improvements observed in problems involving patterns at multiple spatial scales. In the investigation presented in the next section, we draw on insights from the existing literature while also contributing our own strategies and observations.

\section{A case study:  Squeezed kernel functions \label{S4}}

In this section, we illustrate the two kernel learning methods introduced in Section~\ref{S3} using two-dimensional input data ($n = 2$), where each data point $\mathbf{x}$ is encoded in the displacement of a single-mode displaced squeezed vacuum state, as defined in Eq.~(\ref{sd}). With respect to the quantum feature map, the resulting kernel inherits the complex squeezing hyperparameter, which we optimize using the described learning techniques. We refer to this kernel as the \textit{squeezed kernel}. It possesses two key properties: \textit{(i) } it enables intuitive visualization, and \textit{(ii)} it can be estimated experimentally through techniques well-known to the quantum optics community.

To derive the squeezed kernel, we start with the quantum feature map defined by mapping the two-dimensional input $\mathbf{x} = ([\mathbf{x}]_1, [\mathbf{x}]_2)$ to the displacement of a displaced squeezed vacuum state $\ket{\alpha; \zeta}$, as given in Eq.~(\ref{sd}):
\begin{equation}
\mathbf{x}=([\mathbf{x}]_1,[\mathbf{x}]_2)\rightarrow \ket{\frac{1}{\sqrt{2}}([\mathbf{x}]_1+\mathrm{i}~[\mathbf{x}]_2);\zeta} \equiv \ket{\mathbf{x};\zeta},\label{sfm}
\end{equation}
where the squeezing parameter is $\zeta = r e^{i\phi}$.

The corresponding Wigner function is given by
\begin{equation}
W_{\mathbf{x};\zeta}(q,p)=\frac{1}{\pi} e^{-e^{2r}X^2 - e^{-2r}P^2},
\end{equation}
with $X= \cos\phi ~(q-[\mathbf{x}]_1)+ \sin\phi ~(p-[\mathbf{x}]_2)$ and $P= - \sin\phi~ (q-[\mathbf{x}]_1)+ \cos\phi~ (p-[\mathbf{x}]_2)$. Using Eq.~(\ref{KW}) for $N = 1$, this leads to the squeezed kernel:
\begin{equation}
k_{\gamma; \zeta}(\mathbf{x}, \mathbf{x}') = \left|\braket{\mathbf{x};\zeta}{\mathbf{x}';\zeta} \right|^2 = \exp\left(-\frac{\gamma}{2} \left[e^{2r}X^2 + e^{-2r}P^2\right]\right), \label{kz}
\end{equation}
where $\gamma = 1$, and  $X=\cos\phi ~([\mathbf{x}]_1-[\mathbf{x}']_1)+ \sin\phi ~([\mathbf{x}]_2-[\mathbf{x}']_2)$, $P= -\sin\phi ~~([\mathbf{x}]_1-[\mathbf{x}']_1)+ \cos\phi ~([\mathbf{x}]_2-[\mathbf{x}']_2)$.

\begin{figure}[htbp]
\centering\includegraphics[width=14cm]{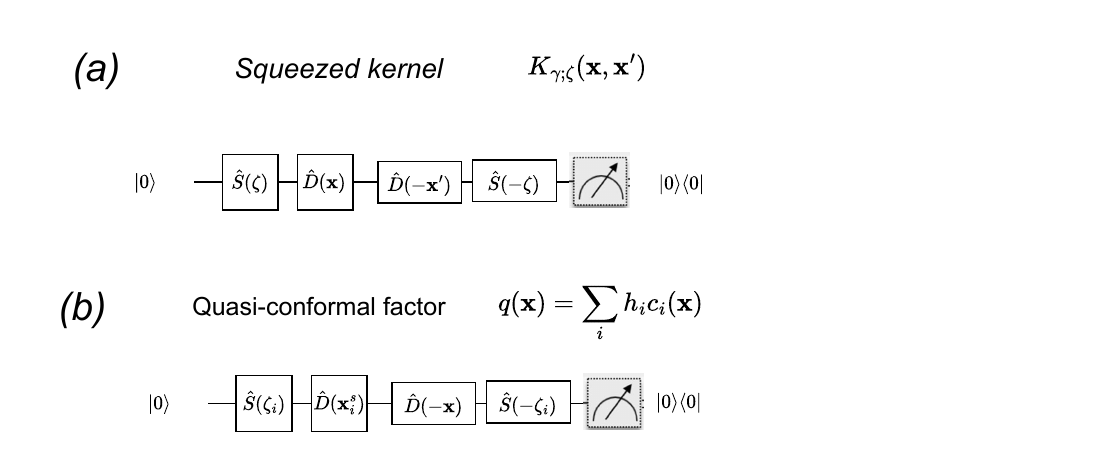}
\caption{\textit{(a)} Circuit for experimentally accessing the squeezed kernel [Eq.~(\ref{kz})], where the probability of measuring zero photons corresponds to $k_{\gamma=1;\zeta}(\mathbf{x},\mathbf{x}')$. \textit{(b)} Add-on circuit for estimating quasi-conformal factors by sampling circuits parameterized by $(\zeta_i,\mathbf{x}_i)$ with probability $h_i$ and measuring zero-photon outcomes, assuming $\sum_i h_i = 1$.\label{fig1}}
\end{figure}

Experimentally, the squeezed kernel can be estimated using the inversion test~\cite{Schuld:2021mml}, illustrated in Fig.~\ref{fig1}~(a). This protocol involves applying a simple sequence of passive optical operations, followed by measuring the probability of detecting zero photons in the output state. It can be shown that this probability directly yields $\left|\braket{\mathbf{x};\zeta}{\mathbf{x}';\zeta} \right|^2$.
Although in practice one does not have straightforward control\footnote{One may repeat the non-commuting blocks in Fig.~\ref{fig1}(a) to partially adjust the variance and squeezing parameters of the kernel.} over the variance of the experimentally produced kernel, this limitation can be circumvented by rescaling the classical data. This allows us to effectively consider squeezed kernels with variable $\gamma$ in our numerical studies.

The squeezed kernel in Eq.(\ref{kz}) defines a family of two-variable Gaussian functions, with the RBF kernel of Eq.(\ref{RBF}) recovered as a special case when $\zeta = 0$. The transformation induced by an RBF kernel is associated with a Euclidean metric, $g_{ij} = \gamma~\delta_{ij}$~\cite{AMARI1999783}, which implies that the transformation preserves the relative arrangement of input data, merely rescaling the feature space uniformly. This uniform rescaling is captured by the magnification factor, given by $\sqrt{\det[g_{ij}]} = \sqrt{G(\mathbf{x})} = \gamma$, which is independent of the data.

For the squeezed kernel, the transformation metric tensor derived from Eq.(\ref{g}) is 
\begin{equation}
    g_{ij}=\gamma\left(\begin{array}{cc}
       \sinh (2 r) \cos (2 \phi )+\cosh (2 r)  & 2 \sinh (2 r) \sin (\phi ) \cos (\phi )  \\
       2 \sinh (2 r) \sin (\phi ) \cos (\phi )  &  \cosh (2 r)-\sinh (2 r) \cos (2 \phi )
    \end{array}\right) \label{me}
\end{equation}
which, like the RBF case, is data-independent and leads to the same magnification factor as the Euclidean metric. However, the squeezed kernel induces anisotropy in the feature space through its squeezing parameters. This is reflected in the metric’s eigenvectors:
\begin{eqnarray}
   \mathbf{e}_1=&(-\tan{\phi},1) \nonumber \\
   \mathbf{e}_2=&(\cot{\phi},1)
\end{eqnarray}
which depend on the squeezing angle $\phi$. These eigenvectors define the principal axes—major and minor, respectively—of the ellipses corresponding to the level sets of Eq.(\ref{kz}). Notably, $\mathbf{e}_1$ also defines  the principal spread direction of the transformation.


In summary, the squeezed kernel used in the examples that follow retains many of the core geometric properties of the RBF kernel—such as data-independent magnification and smooth similarity decay. However, unlike the isotropic RBF kernel, the squeezed kernel introduces a directional bias through the squeezing parameter. This anisotropy allows the kernel to emphasize specific directions in the input space, effectively aligning the similarity measure with the underlying structure of the data. In the next section, we will demonstrate how this directional sensitivity can be guided (learned) to improve performance in certain classification tasks compared to the rotationally symmetric RBF kernel.

\subsection{Learning the squeezing parameter via quantum metric learning}

We illustrate the quantum kernel learning method of Section~\ref{S3} on the co-eccentric circles classification problem, focusing on the optimization of the complex squeezing parameter $\zeta$ via quantum metric learning followed by SVM classification. This allows us to directly compare our approach with a standard RBF kernel SVM and with the original quantum metric learning scheme of \cite{Loyd:1}.

\begin{figure}[htbp]
\centering\includegraphics[width=8
cm]{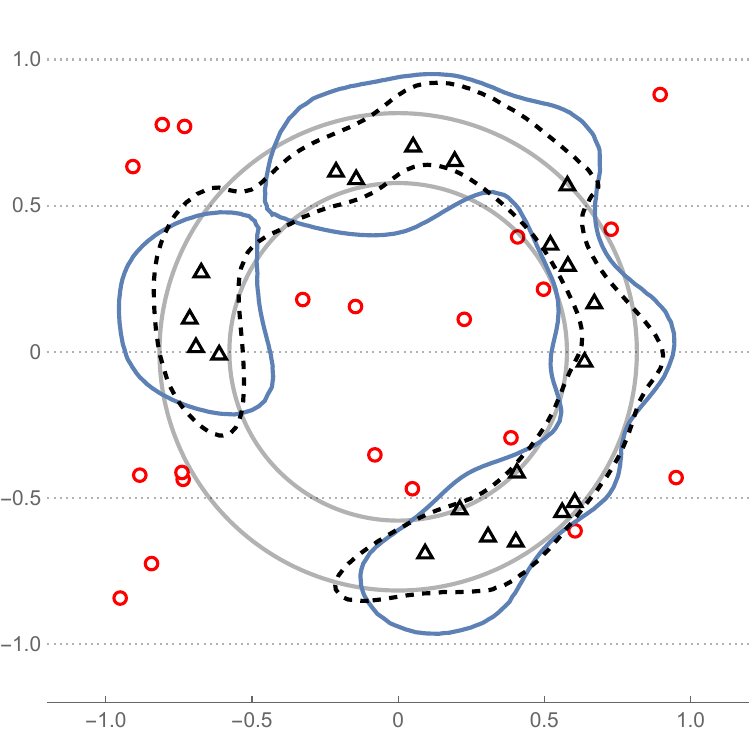}
\caption{Depiction of the co-eccentric circles classification problem with randomly generated training data. Class~A samples are shown as triangles and Class~B samples as circles. The exact decision boundary is indicated by gray circles. The blue solid line corresponds to the decision boundary obtained with the RBF kernel SVM, while the black dashed line shows the boundary obtained with quantum metric learning followed by SVM, i.e., the kernel learning method proposed in this work.\label{fig2}}
\end{figure}

The classification task is defined on a ring centered at the origin with inner radius $r_1=1/\sqrt{3}$ and outer radius $r_2=\sqrt{2/3}$. Class~A ($y=1$) corresponds to data points uniformly sampled inside the ring, while Class~B ($y=-1$) corresponds to points outside the ring. A representative dataset together with the exact decision boundary is shown in Figure~\ref{fig2}. For training, we randomly generate $N_A=N_B=20$ samples per class, while the test set consists of $8$ samples in total. Since all three models achieve nearly perfect training classification (training accuracy consistently above $99\%$), we focus on their predictive power. To estimate average test accuracy, we repeat the experiment with $50$ different random seeds for data generation. The SVM implementation follows \cite{palancz2004svm}, and our code for reproducing all numerical results and figures is available at \cite{Git}.

As a baseline, we first apply the RBF kernel SVM with $\gamma=40$, with results reported in the first row of Table~\ref{tab:1}. The second method is the quantum metric learning scheme of \cite{Loyd:1}, which maximizes the Hilbert--Schmidt distance [Eq.~(\ref{dhs})] between the density matrices [Eq.~(\ref{rhoAB})] associated with the squeezed feature map [Eq.~(\ref{sfm})]. During optimization, the squeezing parameter $\zeta = r e^{i \phi}$ is learned. To reduce the computational cost, we assume $\phi = \arctan([\mathbf{x}]_2/[\mathbf{x}]_1) + \theta$, i.e., the squeezing angle is aligned with the polar angle of the input, up to a tunable offset $\theta$. This ansatz is motivated by the intuition that the minor axis of the squeezing ellipse should be oriented along the radial direction of the data. The decision function for test data is then obtained via the fidelity [Eq.~(\ref{fid})]. Averaged over $50$ seeds, the learned offset is $\theta \approx 0.02 \pm 0.22~\mathrm{rad}$, and the squeezing strength is $0.67 \pm 0.17$.

Finally, our proposed method uses the same quantum metric learning procedure to optimize the squeezing parameter, but instead of relying on fidelity-based classification, we subsequently apply the SVM. This hybrid approach yields the best predictive performance for this task, as summarized in Table~\ref{tab:1}. Figure~\ref{fig2} illustrates, for a representative dataset, the decision boundary obtained by our method compared with that of the RBF kernel SVM. These results highlight the advantage of combining quantum metric learning with SVM method.

\begin{table}[htbp]
\caption{Classification task on co-eccentric circles. The table reports the average test accuracy $\langle A \rangle$ (over $50$ independent runs) together with the corresponding standard deviation, for three methods: (i) RBF kernel SVM, (ii) quantum metric learning  with Hilbert--Schmidt distance, and (iii) QML supplemented with SVM (our proposed method). Both training and test sets are balanced, with $40$ and $8$ data points respectively. For fairness of comparison, the kernel variance for both the RBF and squeezed kernels is set to $\gamma=40$.}
 \label{tab:1}
  \centering
\begin{tabular}{ccc}
\hline
Method & $\left\langle A \right\rangle$ & $\pm \Delta A$ \\
\hline
RBF kernel SVM  & $0.77$ & $0.16$ \\
quantum metric learning & $0.80$ & $0.14$ \\
quantum metric learning  $+$ SVM & $0.82$ & $0.14$ \\

\hline
\end{tabular}
\end{table}

\subsection{`Squeezed' quasi-conformal transformations}

The method of optimizing the kernel via quasi-conformal transformations consists of two runs of the SVM algorithm: the first with an initial kernel function $k(\mathbf{x}, \mathbf{x}')$, which provides a preliminary decision boundary and a set of support vectors, and the second with a modified kernel $\tilde{k}(\mathbf{x}, \mathbf{x}')$ constructed from the support vectors obtained in the first run.

To build intuition, we begin with a simple classification problem where the procedure does not yield an advantage, since the RBF kernel SVM already fits the problem perfectly. Consider a centered circle of radius $r=1/\sqrt{3}$. Class~A ($y=1$) consists of data points randomly distributed in the region between a centered square of side $0.8$ and the circle, while Class~B ($y=-1$) consists of points placed inside the circle.

\begin{figure}[htbp]
\centering
\includegraphics[width=14cm]{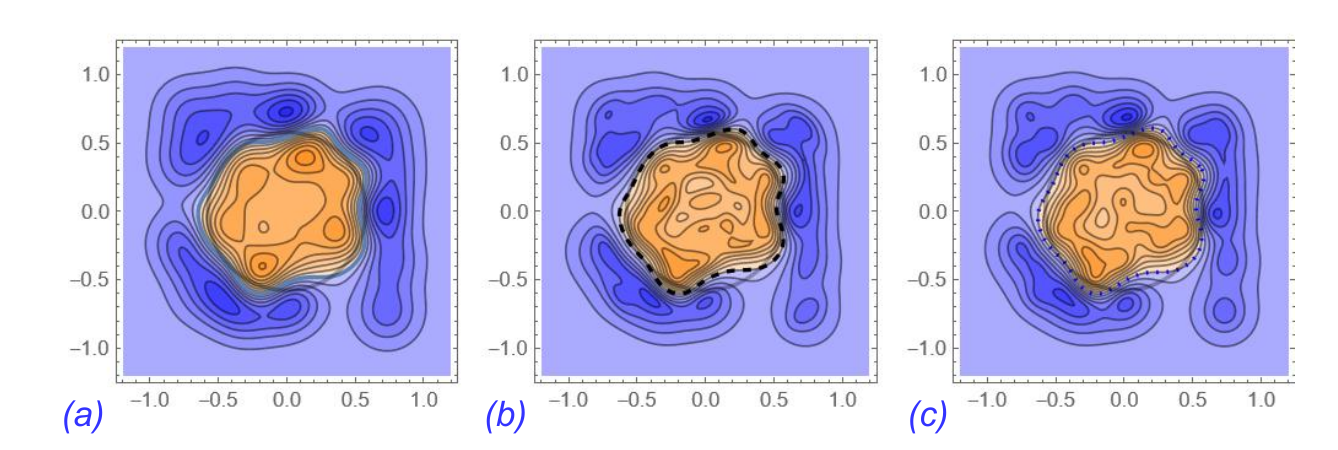}
\caption{Decision functions obtained with three methods: (a) RBF kernel SVM (light blue), (b) quasi-conformal transformation with squeezed kernel (dashed black), and (c) quasi-conformal transformation with RBF kernel (dotted black). The gray circular boundary indicates the real decision function according which the data were generated.
\label{fig3}}
\end{figure}

We follow the procedure below to obtain the decision functions shown in Fig.~\ref{fig3}, using $60$ randomly generated training points:
\begin{itemize}
    \item First, the RBF kernel SVM is applied with kernel $k_{\gamma}(\mathbf{x},\mathbf{x}')$ and $\gamma=40$. This yields $m$ support vectors $\mathbf{x}_j^s$, together with a decision function $f(\mathbf{x})$, where the boundary is given by $f(\mathbf{x})=0$. The result of this step is shown in Fig.~\ref{fig3}(a).
    \item For the second SVM run we introduce the factor
    \begin{equation}
      q(\mathbf{x})=h_0+\sum_{j=1}^m h_j \, k_{\gamma',\zeta_j}(\mathbf{x}^s_j,\mathbf{x}),  \label{sum2}
    \end{equation}
    where $k_{\gamma',\zeta_j}$ is the squeezed kernel defined in Eq.~(\ref{kz}). After testing different prescriptions from the literature \cite{AMARI2002,Qiuze2015}, we adopted the uniform choice $h_j=1$. For the variance we set $\gamma' = 4\gamma$, while the squeezing amplitude was heuristically optimized to $r=0.4$. The modified kernel is then $\tilde{k}(\mathbf{x},\mathbf{x}')=k_{\gamma}(\mathbf{x},\mathbf{x}')\, q(\mathbf{x}) \, q(\mathbf{x}')$, with the squeezing angle $\phi_j$ still to be specified.
    \item To determine $\phi_j$, we use a numerical procedure ensuring that the major axis of the squeezing ellipse is aligned with the preliminary decision boundary from the first run. For each support vector $\mathbf{x}^s_j$, we locate the nearest point $\mathbf{x}^b_j$ on the boundary $f(\mathbf{x})=0$ and compute the gradient $\nabla f(\mathbf{x})|_{\mathbf{x}=\mathbf{x}^b_j}$, whose direction defines $\phi_j$. The resulting decision function and boundary are shown in Fig.~\ref{fig3}(b).
    \item Finally, we apply the same quasi-conformal transformation procedure with $\zeta=0$, i.e., using $q(\mathbf{x})$ built from RBF kernels with $\gamma'=4\gamma$. The corresponding result is shown in Fig.~\ref{fig3}(c).
\end{itemize}

For an experimental realization of the factors $q(\mathbf{x})$ in Eq.~(\ref{sum2}), we propose the quantum optical circuit of Fig.~\ref{fig1}(b). The required superposition of kernel functions $k_{\gamma',\zeta_j}(\mathbf{x}^s_j,\mathbf{x})$ can be implemented by sampling the index $j$, which determines the circuit parameters $(\zeta_j,\mathbf{x}^s_j)$, with probability set by the weights $h_j$.

As expected for this simple problem, all methods achieve nearly identical accuracy on both training and test data. However, Fig.~\ref{fig3} clearly illustrates how the regions around the support vectors are locally modified under quasi-conformal transformations, directly revealing the induced magnification factors. This sharpening of the decision function may not improve performance in this toy example, but it becomes advantageous in more intricate classification tasks, such as the one presented next.

\vspace{1cm}

We now turn to the more intricate task of classifying data separated by a \textit{hypotrochoid} boundary, parametrically defined as
\begin{eqnarray}
    x(\theta) &= (R-r)\cos{\theta} + d \cos{\tfrac{R-r}{r}\theta}, \nonumber \\
    y(\theta) &= (R-r)\sin{\theta} - d \sin{\tfrac{R-r}{r}\theta}, \label{troc}
\end{eqnarray}
with $r=d$, $R/r=1.5$, and $\theta \in [0,2\pi)$. Class~A consists of data randomly generated inside the curve characterized by $R=0.15$, $r=d=0.1$, while Class~B consists of data outside this curve but inside the larger one defined by $R=0.36$, $r=d=0.24$. The training set is balanced and contains $40$ points, while the test set contains $80$ points.

We apply the same procedure as described in the previous example, but with one modification in the third step, namely the determination of the squeezing angle $\phi_j$. Owing to the large number of samples required for statistical accuracy, we did not incorporate an optimization loop for $\phi_j$. Instead, we used the formula
\begin{equation}
\phi = \arctan\!\left(\frac{[\mathbf{x}]_2}{[\mathbf{x}]_1}\right) + \frac{\pi}{2},
\end{equation}
where $([\mathbf{x}]_1,[\mathbf{x}]_2)$ are the coordinates of the corresponding support vector. The offset angle $\pi/2$ was chosen empirically.

\begin{figure}[htbp]
\centering\includegraphics[width=10
cm]{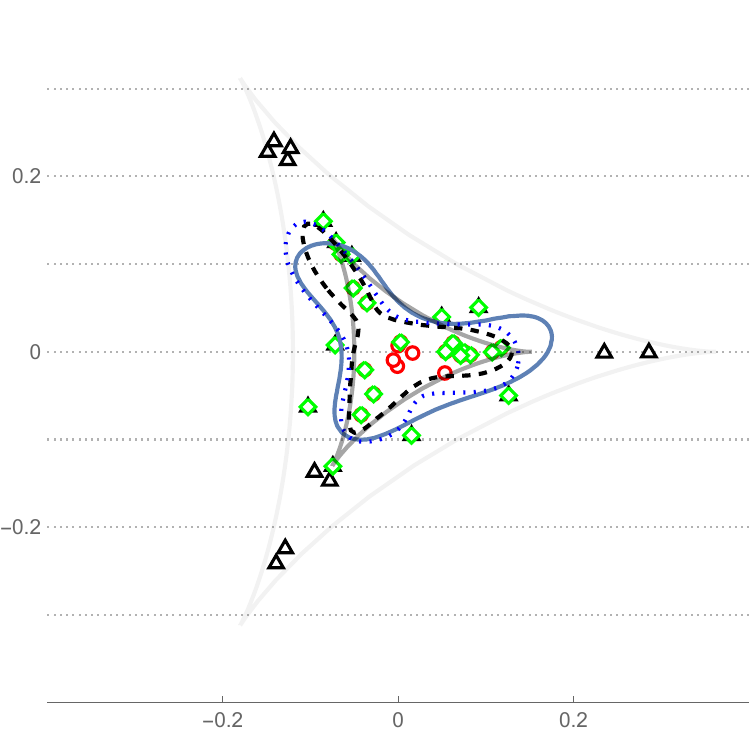}
\caption{Classification task with hypotrochoid separating boundaries (light gray lines). Class~A data are shown as red circles, Class~B as black triangles, and support vectors from the first RBF kernel SVM run as green diamonds. Decision boundaries are indicated as follows: blue solid line (RBF kernel SVM), black dashed line (squeezed quasi-conformal transformation), and blue dotted line (RBF quasi-conformal transformation). The kernel variance is set to $\gamma = 2\times 10^{2}$.
 \label{fig4}}
\end{figure}

We tested the procedure by measuring the accuracy on the test set, and the results are summarized in Table~\ref{tab:2}. Each entry represents an average accuracy over $100$ runs with different random seeds for generating the training and test data. The results are reported for two different values of $\gamma$ while the rest of parameters in Eq.(\ref{sum2}) are set as: $\gamma' = 4\gamma$, $r=0.5$, $h_0=0$, $h_{i\ne0}=1$. Importantly, quasi-conformal transformations introduce more than a mere rescaling of the kernel function: the induced magnification in specific regions reveals fine details of smaller patterns embedded in a larger structure. An example of decision boundaries obtained by the three methods is shown in Fig.~\ref{fig4}.

\begin{table}[htbp]
\caption{Classification task with hypotrochoid separating boundaries (see Fig.~\ref{fig4}). Reported is the average test accuracy achieved by: \textit{(a)} RBF kernel SVM, \textit{(b)} RBF kernel SVM followed by squeezed quasi-conformal transformations, and \textit{(c)} RBF kernel SVM followed by RBF quasi-conformal transformations. The training set contains $40$ points ($20$ per class), and the test set contains $80$ randomly generated points. Statistics are averaged over $100$ runs. For the squeezed quasi-conformal transformation, the factor $q(\mathbf{x})$ is given by Eq.~(\ref{sum2}) with $h_0=0$, $r=0.5$, and $\phi_j=\arctan([\mathbf{x}_j]_2/[\mathbf{x}_j]_1)+\pi/2$. For the RBF case, Eq.~(\ref{sum2}) is used with $r=0$.}
  \label{tab:2}
  \centering
\begin{tabular}{cccc}

\hline
Method & $\left\langle A \right\rangle$ $\pm \Delta A$  &   \\
\hline
$\gamma$  & $10^2$ & $2 \times 10^2$ \\
\hline
 RBF SVM & $0.85\pm0.03$ & $0.86\pm 0.04$  \\
RBF SVM +  Squeezed CT SVM & $0.89\pm 0.05$ & $0.89\pm 0.05$    \\
RBF SVM +  RBF CT SVM & $0.87\pm 0.04$ & $0.88 \pm 0.05$  \\
\hline
\end{tabular}
\end{table}

\section{Conclusions}

In this work we have translated two classical kernel learning methods into the framework of quantum circuits. We showed that the Fisher criterion is closely related to a known technique in the quantum literature, namely quantum kernel learning with Hilbert--Schmidt distance \cite{Loyd:1}, and that the associated methods can be naturally employed as quantum kernel learning stages prior to the application of SVM. For the second method, based on quasi-conformal transformations of kernels, we demonstrated how such transformations can be realized with quantum optical circuits. Through our examples we also highlighted the effectiveness of single-mode displaced squeezed states in constructing useful kernels.

The methods presented here can be directly extended to higher-dimensional quantum circuits, offering kernels suited to more intricate classification tasks. The topic of quasi-conformal transformations with quantum circuits in particular deserves further study: beyond the circuit constructions suggested here for implementing the quasi-conformal factors, alternative strategies may be possible, potentially involving entanglement. An important open question is whether full conformal transformations can be achieved with parametrized quantum circuits.

Finally, we note that all examples in this work involved Gaussian quantum operations. We also carried out preliminary tests (results not reported) with non-Gaussian states, specifically displaced squeezed single photons, but observed no advantage in the resulting kernels. Nevertheless, the landscape of non-Gaussian quantum operations is vast, and we expect that genuinely non-classical (i.e., non-Gaussian) resources, when combined with the methods proposed here, may lead to outcomes beyond the reach of classical computation.

\section*{Acknowledgments}
This work was supported by   the European Union’s Horizon
Europe research and innovation program under grant agreement No.101092766 (ALLEGRO Project). AM is thankful to Ralntion Komini for helpful discussions.

\bibliography{references}

\begin{thebibliography}{28}
\expandafter\ifx\csname natexlab\endcsname\relax\def\natexlab#1{#1}\fi
\expandafter\ifx\csname bibnamefont\endcsname\relax
  \def\bibnamefont#1{#1}\fi
\expandafter\ifx\csname bibfnamefont\endcsname\relax
  \def\bibfnamefont#1{#1}\fi
\expandafter\ifx\csname citenamefont\endcsname\relax
  \def\citenamefont#1{#1}\fi
\expandafter\ifx\csname url\endcsname\relax
  \def\url#1{\texttt{#1}}\fi
\expandafter\ifx\csname urlprefix\endcsname\relax\def\urlprefix{URL }\fi
\providecommand{\bibinfo}[2]{#2}
\providecommand{\eprint}[2][]{\url{#2}}

\bibitem[{\citenamefont{Vapnik and Chervonenkis}(1964)}]{Vapnik:1}
\bibinfo{author}{\bibfnamefont{V.}~\bibnamefont{Vapnik}} \bibnamefont{and} \bibinfo{author}{\bibfnamefont{A.}~\bibnamefont{Chervonenkis}}, \bibinfo{journal}{Automation and Remote Control} p.~\bibinfo{pages}{25} (\bibinfo{year}{1964}).

\bibitem[{\citenamefont{Cortes and Vapnik}(1995)}]{Vapnik:2}
\bibinfo{author}{\bibfnamefont{C.}~\bibnamefont{Cortes}} \bibnamefont{and} \bibinfo{author}{\bibfnamefont{V.}~\bibnamefont{Vapnik}}, \bibinfo{journal}{Machine Learning} \textbf{\bibinfo{volume}{20}}, \bibinfo{pages}{273} (\bibinfo{year}{1995}), \urlprefix\url{https://doi.org/10.1007/BF00994018}.

\bibitem[{\citenamefont{Schuld and Killoran}(2019)}]{Schuld:1}
\bibinfo{author}{\bibfnamefont{M.}~\bibnamefont{Schuld}} \bibnamefont{and} \bibinfo{author}{\bibfnamefont{N.}~\bibnamefont{Killoran}}, \bibinfo{journal}{Phys. Rev. Lett.} \textbf{\bibinfo{volume}{122}}, \bibinfo{pages}{040504} (\bibinfo{year}{2019}), \urlprefix\url{https://link.aps.org/doi/10.1103/PhysRevLett.122.040504}.

\bibitem[{\citenamefont{Havl{\'i}{\v{c}}ek et~al.}(2019)\citenamefont{Havl{\'i}{\v{c}}ek, C{\'o}rcoles, Temme, Harrow, Kandala, Chow, and Gambetta}}]{Harrow:1}
\bibinfo{author}{\bibfnamefont{V.}~\bibnamefont{Havl{\'i}{\v{c}}ek}}, \bibinfo{author}{\bibfnamefont{A.~D.} \bibnamefont{C{\'o}rcoles}}, \bibinfo{author}{\bibfnamefont{K.}~\bibnamefont{Temme}}, \bibinfo{author}{\bibfnamefont{A.~W.} \bibnamefont{Harrow}}, \bibinfo{author}{\bibfnamefont{A.}~\bibnamefont{Kandala}}, \bibinfo{author}{\bibfnamefont{J.~M.} \bibnamefont{Chow}}, \bibnamefont{and} \bibinfo{author}{\bibfnamefont{J.~M.} \bibnamefont{Gambetta}}, \bibinfo{journal}{Nature} \textbf{\bibinfo{volume}{567}}, \bibinfo{pages}{209} (\bibinfo{year}{2019}), ISSN \bibinfo{issn}{1476-4687}, \urlprefix\url{https://doi.org/10.1038/s41586-019-0980-2}.

\bibitem[{\citenamefont{Thanasilp et~al.}(2024)\citenamefont{Thanasilp, Wang, Cerezo, and Holmes}}]{Thanasilp}
\bibinfo{author}{\bibfnamefont{S.}~\bibnamefont{Thanasilp}}, \bibinfo{author}{\bibfnamefont{S.}~\bibnamefont{Wang}}, \bibinfo{author}{\bibfnamefont{M.}~\bibnamefont{Cerezo}}, \bibnamefont{and} \bibinfo{author}{\bibfnamefont{Z.}~\bibnamefont{Holmes}}, \bibinfo{journal}{Nature Communications} \textbf{\bibinfo{volume}{15}}, \bibinfo{pages}{5200} (\bibinfo{year}{2024}), \urlprefix\url{https://doi.org/10.1038/s41467-024-49287-w}.

\bibitem[{\citenamefont{Kübler et~al.}(2021)\citenamefont{Kübler, Buchholz, and Schölkopf}}]{Kubler}
\bibinfo{author}{\bibfnamefont{J.~M.} \bibnamefont{Kübler}}, \bibinfo{author}{\bibfnamefont{S.}~\bibnamefont{Buchholz}}, \bibnamefont{and} \bibinfo{author}{\bibfnamefont{B.}~\bibnamefont{Schölkopf}}, in \emph{\bibinfo{booktitle}{35th Conference on Neural Information Processing Systems}} (\bibinfo{year}{2021}).

\bibitem[{\citenamefont{Shaydulin and Wild}(2022)}]{Ruslan}
\bibinfo{author}{\bibfnamefont{R.}~\bibnamefont{Shaydulin}} \bibnamefont{and} \bibinfo{author}{\bibfnamefont{S.~M.} \bibnamefont{Wild}}, \bibinfo{journal}{Phys. Rev. A} \textbf{\bibinfo{volume}{106}}, \bibinfo{pages}{042407} (\bibinfo{year}{2022}), \urlprefix\url{https://link.aps.org/doi/10.1103/PhysRevA.106.042407}.

\bibitem[{\citenamefont{Flórez-Ablan et~al.}(2025)\citenamefont{Flórez-Ablan, Roth, and Schnabel}}]{Florez}
\bibinfo{author}{\bibfnamefont{R.}~\bibnamefont{Flórez-Ablan}}, \bibinfo{author}{\bibfnamefont{M.}~\bibnamefont{Roth}}, \bibnamefont{and} \bibinfo{author}{\bibfnamefont{J.}~\bibnamefont{Schnabel}}, \bibinfo{journal}{Quantum Science and Technology} \textbf{\bibinfo{volume}{10}}, \bibinfo{pages}{035051} (\bibinfo{year}{2025}).

\bibitem[{\citenamefont{Liu et~al.}(2021)\citenamefont{Liu, Arunachalam, and Temme}}]{Liu}
\bibinfo{author}{\bibfnamefont{Y.}~\bibnamefont{Liu}}, \bibinfo{author}{\bibfnamefont{S.}~\bibnamefont{Arunachalam}}, \bibnamefont{and} \bibinfo{author}{\bibfnamefont{K.}~\bibnamefont{Temme}}, \bibinfo{journal}{Nature Physics} \textbf{\bibinfo{volume}{17}}, \bibinfo{pages}{1013} (\bibinfo{year}{2021}), \urlprefix\url{https://doi.org/10.1038/s41567-021-01287-z}.

\bibitem[{\citenamefont{Huang et~al.}(2021)\citenamefont{Huang, Broughton, Mohseni, Babbush, Boixo, Neven, and McClean}}]{Huang}
\bibinfo{author}{\bibfnamefont{H.-Y.} \bibnamefont{Huang}}, \bibinfo{author}{\bibfnamefont{M.}~\bibnamefont{Broughton}}, \bibinfo{author}{\bibfnamefont{M.}~\bibnamefont{Mohseni}}, \bibinfo{author}{\bibfnamefont{R.}~\bibnamefont{Babbush}}, \bibinfo{author}{\bibfnamefont{S.}~\bibnamefont{Boixo}}, \bibinfo{author}{\bibfnamefont{H.}~\bibnamefont{Neven}}, \bibnamefont{and} \bibinfo{author}{\bibfnamefont{J.~R.} \bibnamefont{McClean}}, \bibinfo{journal}{Nature Communications} \textbf{\bibinfo{volume}{12}}, \bibinfo{pages}{2631} (\bibinfo{year}{2021}), \urlprefix\url{https://doi.org/10.1038/s41467-021-22539-9}.

\bibitem[{\citenamefont{J{\"a}ger and Krems}(2023)}]{Jager}
\bibinfo{author}{\bibfnamefont{J.}~\bibnamefont{J{\"a}ger}} \bibnamefont{and} \bibinfo{author}{\bibfnamefont{R.~V.} \bibnamefont{Krems}}, \bibinfo{journal}{Nature Communications} \textbf{\bibinfo{volume}{14}}, \bibinfo{pages}{576} (\bibinfo{year}{2023}), ISSN \bibinfo{issn}{2041-1723}, \urlprefix\url{https://doi.org/10.1038/s41467-023-36144-5}.

\bibitem[{\citenamefont{Schölkopf and Smola}(2001)}]{Scholkopf}
\bibinfo{author}{\bibfnamefont{B.}~\bibnamefont{Schölkopf}} \bibnamefont{and} \bibinfo{author}{\bibfnamefont{A.~J.} \bibnamefont{Smola}}, \emph{\bibinfo{title}{Learning with Kernels: Support Vector Machines, Regularization, Optimization, and Beyond}} (\bibinfo{publisher}{{The MIT Press}}, \bibinfo{year}{2001}), ISBN \bibinfo{isbn}{9780262256933}.

\bibitem[{\citenamefont{Schuld}(2021)}]{schuld:2}
\bibinfo{author}{\bibfnamefont{M.}~\bibnamefont{Schuld}}, \bibinfo{journal}{arXiv preprint arXiv:2101.11020}  (\bibinfo{year}{2021}).

\bibitem[{\citenamefont{Hubregtsen et~al.}(2022)\citenamefont{Hubregtsen, Wierichs, Gil-Fuster, Derks, Faehrmann, and Meyer}}]{Hubregtsen}
\bibinfo{author}{\bibfnamefont{T.}~\bibnamefont{Hubregtsen}}, \bibinfo{author}{\bibfnamefont{D.}~\bibnamefont{Wierichs}}, \bibinfo{author}{\bibfnamefont{E.}~\bibnamefont{Gil-Fuster}}, \bibinfo{author}{\bibfnamefont{P.-J. H.~S.} \bibnamefont{Derks}}, \bibinfo{author}{\bibfnamefont{P.~K.} \bibnamefont{Faehrmann}}, \bibnamefont{and} \bibinfo{author}{\bibfnamefont{J.~J.} \bibnamefont{Meyer}}, \bibinfo{journal}{Phys. Rev. A} \textbf{\bibinfo{volume}{106}}, \bibinfo{pages}{042431} (\bibinfo{year}{2022}), \urlprefix\url{https://link.aps.org/doi/10.1103/PhysRevA.106.042431}.

\bibitem[{\citenamefont{Glick et~al.}(2024)\citenamefont{Glick, , Gujarati, Córcoles, Kim, Kandala, Gambetta, and Temme}}]{Glick}
\bibinfo{author}{\bibfnamefont{J.~R.} \bibnamefont{Glick}}, , \bibinfo{author}{\bibfnamefont{T.~P.} \bibnamefont{Gujarati}}, \bibinfo{author}{\bibfnamefont{A.~D.} \bibnamefont{Córcoles}}, \bibinfo{author}{\bibfnamefont{Y.}~\bibnamefont{Kim}}, \bibinfo{author}{\bibfnamefont{A.}~\bibnamefont{Kandala}}, \bibinfo{author}{\bibfnamefont{J.~M.} \bibnamefont{Gambetta}}, \bibnamefont{and} \bibinfo{author}{\bibfnamefont{K.}~\bibnamefont{Temme}}, \bibinfo{journal}{Nature Physics} \textbf{\bibinfo{volume}{20}}, \bibinfo{pages}{479} (\bibinfo{year}{2024}), \urlprefix\url{https://doi.org/10.1038/s41567-023-02340-9}.

\bibitem[{\citenamefont{Lloyd et~al.}(2020)\citenamefont{Lloyd, Schuld, Ijaz, Izaac, and Killoran}}]{Loyd:1}
\bibinfo{author}{\bibfnamefont{S.}~\bibnamefont{Lloyd}}, \bibinfo{author}{\bibfnamefont{M.}~\bibnamefont{Schuld}}, \bibinfo{author}{\bibfnamefont{A.}~\bibnamefont{Ijaz}}, \bibinfo{author}{\bibfnamefont{J.}~\bibnamefont{Izaac}}, \bibnamefont{and} \bibinfo{author}{\bibfnamefont{N.}~\bibnamefont{Killoran}}, \emph{\bibinfo{title}{Quantum embeddings for machine learning}} (\bibinfo{year}{2020}), \eprint{2001.03622}, \urlprefix\url{https://arxiv.org/abs/2001.03622}.

\bibitem[{\citenamefont{Amari and Wu}(1999)}]{AMARI1999783}
\bibinfo{author}{\bibfnamefont{S.}~\bibnamefont{Amari}} \bibnamefont{and} \bibinfo{author}{\bibfnamefont{S.}~\bibnamefont{Wu}}, \bibinfo{journal}{Neural Networks} \textbf{\bibinfo{volume}{12}}, \bibinfo{pages}{783} (\bibinfo{year}{1999}), ISSN \bibinfo{issn}{0893-6080}, \urlprefix\url{https://www.sciencedirect.com/science/article/pii/S0893608099000325}.

\bibitem[{\citenamefont{Wu and Amari}(2002)}]{AMARI2002}
\bibinfo{author}{\bibfnamefont{S.}~\bibnamefont{Wu}} \bibnamefont{and} \bibinfo{author}{\bibfnamefont{S.}~\bibnamefont{Amari}}, \bibinfo{journal}{Neural Processing Letters} \textbf{\bibinfo{volume}{15}}, \bibinfo{pages}{59} (\bibinfo{year}{2002}).

\bibitem[{\citenamefont{Williams et~al.}(2007)\citenamefont{Williams, Li, Feng, and Wu}}]{Wu2007}
\bibinfo{author}{\bibfnamefont{P.}~\bibnamefont{Williams}}, \bibinfo{author}{\bibfnamefont{S.}~\bibnamefont{Li}}, \bibinfo{author}{\bibfnamefont{J.}~\bibnamefont{Feng}}, \bibnamefont{and} \bibinfo{author}{\bibfnamefont{S.}~\bibnamefont{Wu}}, \bibinfo{journal}{IEEE Transactions on Neural Networks} \textbf{\bibinfo{volume}{18}}, \bibinfo{pages}{942} (\bibinfo{year}{2007}).

\bibitem[{\citenamefont{Skotinioti}(2024)}]{Tth}
\bibinfo{author}{\bibfnamefont{T.~I.} \bibnamefont{Skotinioti}}, \emph{\bibinfo{title}{B{S}c thesis: Improving the performance of quantum kernel support vector classifiers via conformal transformations, {N}ational and {K}apodistrian {U}niversity of {A}thens}} (\bibinfo{year}{2024}).

\bibitem[{\citenamefont{Schuld and Petruccione}(2021)}]{Schuld:2021mml}
\bibinfo{author}{\bibfnamefont{M.}~\bibnamefont{Schuld}} \bibnamefont{and} \bibinfo{author}{\bibfnamefont{F.}~\bibnamefont{Petruccione}}, \emph{\bibinfo{title}{{Machine Learning with Quantum Computers}}}, Quantum Science and Technology (\bibinfo{publisher}{Springer}, \bibinfo{address}{Cham}, \bibinfo{year}{2021}), ISBN \bibinfo{isbn}{978-3-030-83097-7, 978-3-030-83100-4, 978-3-030-83098-4}.

\bibitem[{\citenamefont{Schleich}(2001)}]{Schleich}
\bibinfo{author}{\bibfnamefont{W.~P.} \bibnamefont{Schleich}}, \emph{\bibinfo{title}{Quantum Optics in Phase Space}} (\bibinfo{publisher}{Wiley‐VCH Verlag Berlin GmbH}, \bibinfo{year}{2001}), ISBN \bibinfo{isbn}{9783527294350}.

\bibitem[{\citenamefont{Li et~al.}(2022)\citenamefont{Li, Zhang, and Wang}}]{LI2022128088}
\bibinfo{author}{\bibfnamefont{L.~H.} \bibnamefont{Li}}, \bibinfo{author}{\bibfnamefont{D.-B.} \bibnamefont{Zhang}}, \bibnamefont{and} \bibinfo{author}{\bibfnamefont{Z.}~\bibnamefont{Wang}}, \bibinfo{journal}{Physics Letters A} \textbf{\bibinfo{volume}{436}}, \bibinfo{pages}{128088} (\bibinfo{year}{2022}), ISSN \bibinfo{issn}{0375-9601}, \urlprefix\url{https://www.sciencedirect.com/science/article/pii/S0375960122001700}.

\bibitem[{\citenamefont{Mehta and Roy}(2024)}]{MEHTA2024129704}
\bibinfo{author}{\bibfnamefont{V.}~\bibnamefont{Mehta}} \bibnamefont{and} \bibinfo{author}{\bibfnamefont{U.}~\bibnamefont{Roy}}, \bibinfo{journal}{Physics Letters A} \textbf{\bibinfo{volume}{519}}, \bibinfo{pages}{129704} (\bibinfo{year}{2024}), ISSN \bibinfo{issn}{0375-9601}, \urlprefix\url{https://www.sciencedirect.com/science/article/pii/S0375960124003980}.

\bibitem[{\citenamefont{Komini}(2025)}]{Rth}
\bibinfo{author}{\bibfnamefont{R.}~\bibnamefont{Komini}}, \emph{\bibinfo{title}{B{S}c thesis: Minimizing the quantum resources for a quantum distance classifier, {N}ational and {K}apodistrian {U}niversity of {A}thens}} (\bibinfo{year}{2025}).

\bibitem[{\citenamefont{Xiong et~al.}(2017)\citenamefont{Xiong, Yu, Yang, Swamy, and Yu}}]{Qiuze2015}
\bibinfo{author}{\bibfnamefont{H.}~\bibnamefont{Xiong}}, \bibinfo{author}{\bibfnamefont{W.}~\bibnamefont{Yu}}, \bibinfo{author}{\bibfnamefont{X.}~\bibnamefont{Yang}}, \bibinfo{author}{\bibfnamefont{M.~N.~S.} \bibnamefont{Swamy}}, \bibnamefont{and} \bibinfo{author}{\bibfnamefont{Q.}~\bibnamefont{Yu}}, \bibinfo{journal}{IEEE Transactions on Neural Networks and Learning Systems} \textbf{\bibinfo{volume}{28}}, \bibinfo{pages}{149} (\bibinfo{year}{2017}).

\bibitem[{\citenamefont{Pal{\'a}ncz and Volgyesi}(2004)}]{palancz2004svm}
\bibinfo{author}{\bibfnamefont{B.}~\bibnamefont{Pal{\'a}ncz}} \bibnamefont{and} \bibinfo{author}{\bibfnamefont{L.}~\bibnamefont{Volgyesi}}, \emph{\bibinfo{title}{Support vector classifier via mathematica}}, \bibinfo{howpublished}{Wolfram Research Mathematica Information Center} (\bibinfo{year}{2004}), \bibinfo{note}{e-publication, accessed: 2025-09-10}, \urlprefix\url{http://library.wolfram.com/infocenter/MathSource/5293/}.

\bibitem[{\citenamefont{Mandilara}(2025)}]{Git}
\bibinfo{author}{\bibfnamefont{A.}~\bibnamefont{Mandilara}}, \emph{\bibinfo{title}{Learning-kernels}} (\bibinfo{year}{2025}), \urlprefix\url{https://github.com/optcomm/learning-kernels.git}.

\end{thebibliography}

\end{document}